\documentclass[reprint,aip,apl,twocolumn]{revtex4-1}
\usepackage{graphicx}
\usepackage{color}
\usepackage{dcolumn}
\usepackage{enumitem}
\usepackage{epstopdf}

\begin{document}

\preprint{AIP/123-QED}

\title{Double Fe-impurity charge state in the topological insulator Bi$_2$Se$_3$}

\author{V.S. Stolyarov}
\email{vasiliy@travel.ru}
\affiliation{Laboratoire de Physique et d'Etudes des Materiaux, ESPCI-Paris, CNRS and UPMC Univ Paris 6 - UMR 8213, 10 rue Vauquelin, 75005 Paris, France}
\affiliation{Moscow Institute of Physics and Technology, 141700 Dolgoprudny,  Russia}
\affiliation{Institute of  Solid State Physics, Russian Academy of Sciences, 142432 Chernogolovka, Russia}
\affiliation{Moscow State University, Leninskie Gory, 1/3, 119992 Moscow, Russia}
\affiliation{National University of Science and Technology MISIS, 4 Leninsky prosp., 119049 Moscow, Russia}
\affiliation{Solid State Physics Department, KFU, 420008 Kazan, Russia}
\author{S.V. Remizov}
\affiliation{Dukhov Research Institute of Automatics (VNIIA), 127055 Moscow, Russia}
\affiliation{Kotel'nikov Institute of Radio-engineering and Electronics, Russian Academy of Sciences, 125009 Moscow, Russia}
\author{D.S. Shapiro}
\affiliation{Dukhov Research Institute of Automatics (VNIIA), 127055 Moscow, Russia}
\affiliation{Kotel'nikov Institute of Radio-engineering and Electronics, Russian Academy of Sciences, 125009 Moscow, Russia}
\affiliation{Moscow Institute of Physics and Technology, 141700 Dolgoprudny,  Russia}
\author{S. Pons}
\affiliation{Laboratoire de Physique et d'Etudes des Materiaux, ESPCI-Paris, CNRS and UPMC Univ Paris 6 - UMR 8213, 10 rue Vauquelin, 75005 Paris, France}
\author{S. Vlaic}
\affiliation{Laboratoire de Physique et d'Etudes des Materiaux, ESPCI-Paris, CNRS and UPMC Univ Paris 6 - UMR 8213, 10 rue Vauquelin, 75005 Paris, France}
\author{H. Aubin}
\affiliation{Laboratoire de Physique et d'Etudes des Materiaux, ESPCI-Paris, CNRS and UPMC Univ Paris 6 - UMR 8213, 10 rue Vauquelin, 75005 Paris, France}
\author{D.S. Baranov}
\affiliation{Moscow Institute of Physics and Technology, 141700 Dolgoprudny,  Russia}
\affiliation{Institute of  Solid State Physics, Russian Academy of Sciences, 142432 Chernogolovka, Russia}
\affiliation{Institut des Nanosciences de Paris, Sorbonne Universites, UPMC Univ Paris 6 and CNRS-UMR 7588, F-75005 Paris, France}

\author{Ch. Brun}
\affiliation{Institut des Nanosciences de Paris, Sorbonne Universites, UPMC Univ Paris 6 and CNRS-UMR 7588, F-75005 Paris, France}

\author{L.V. Yashina}
\affiliation{Moscow State University, Leninskie Gory, 1/3, 119992 Moscow, Russia}

\author{S.I. Bozhko}
\affiliation{Institute of  Solid State Physics, Russian Academy of Sciences, 142432 Chernogolovka, Russia}

\author{T. Cren}
\affiliation{Institut des Nanosciences de Paris, Sorbonne Universites, UPMC Univ Paris 6 and CNRS-UMR 7588, F-75005 Paris, France}

\author{W.V. Pogosov}
\affiliation{Dukhov Research Institute of Automatics (VNIIA), 127055 Moscow, Russia}
\affiliation{Institute of Theoretical and Applied Electrodynamics, Russian Academy of Sciences, 125412 Moscow, Russia}
\affiliation{Moscow Institute of Physics and Technology, 141700 Dolgoprudny,  Russia}

\author{D. Roditchev}
\affiliation{Laboratoire de Physique et d'Etudes des Materiaux, ESPCI-Paris, CNRS and UPMC Univ Paris 6 - UMR 8213, 10 rue Vauquelin, 75005 Paris, France}
\affiliation{Institut des Nanosciences de Paris, Sorbonne Universites, UPMC Univ Paris 6 and CNRS-UMR 7588, F-75005 Paris, France}
\affiliation{Moscow Institute of Physics and Technology, 141700 Dolgoprudny, Russia}

\date{\today}

\begin{abstract}
The influence of individual impurities of Fe on the electronic properties of topological insulator Bi$_2$Se$_3$ is studied by Scanning Tunneling Microscopy. The microscope tip is used in order to remotely charge/discharge Fe impurities. The charging process is shown to depend on the impurity location in the crystallographic unit cell, on the presence  of other Fe impurities in the close vicinity, as well as on the overall doping level of the crystal.  We present a qualitative explanation of the observed phenomena in terms of tip-induced local band bending. Our observations evidence that the specific impurity neighborhood and the position of the Fermi energy with respect to the Dirac point and bulk bands have both to be taken into account when considering the electron scattering on the disorder in topological insulators.
\end{abstract}


\maketitle

The uniqueness of the electronic properties of topological insulators (TIs) and specifically the topological protection of conduction electrons in these materials make them interesting for applications in quantum electronics \cite{hasan,qi,fu98,moore,hsieh}. The understanding of the impurity scattering effects on the properties of TIs is therefore of primary importance \cite{chen,cui,polyakov,yee,eelbo}. In this context, the role of magnetic impurities is still unclear as their presence may lead to rather complicated phenomena \cite{xu}. For instance, magnetic impurities in TIs can support long-range magnetic order \cite{chen,wray,hor}, open an energy gap at the Dirac point, and even result in formation of a quantum anomalous Hall state \cite{yu}. The possibility to produce a long-range ordering of impurity spins could be of a great interest for realization of novel electronic states, quantum computing, spintronics \cite{song}.

The charge screening being rather poor in Dirac materials, individual atomic impurities are likely to be charged. This charging could have a dramatic impact on the efficiency of novel field-effect nano-devices \cite{xiu} based on topological insulators. Charged individual Fe-impurities embedded in Bi$_2$Se$_3$ have already been reported in previous works \cite{song,west,schlenk} and produce ring-shaped spectral features in conductance images of scanning tunneling microscopy experiments. Also, the observation of impurity charging is not exclusive to TIs: ionization rings were observed near defects in several  materials such as semiconductors \cite{gupta,teichmann,wijnheijmer,marczinowski}, graphene \cite{graphene} and semiconductor surfaces with deposited Co-islands \cite{muzychenko}. In our work, we show that the position of the Fermi energy with respect to the Dirac point and the bulk bands\cite{Kapustin} plays a crucial role in the charge screening and scattering in TIs. We also evidence the strong influence of neighboring impurities in the charging of an individual iron atom.

Our experiments were done by means of 3 complementary experimental techniques. Scanning Tunneling Microscopy/Spectroscopy (STM/STS), Atomic Force Microscopy (AFM) and Angle Resolved Photoemission Spectroscopy (ARPES) were used for the study of Bi$_2$Se$_3$ single crystals in which 0.2\% of Fe-impurities were incorporated.
The studied samples were cleaved \emph{in situ} under ultra-high vacuum. This procedure enables the production of atomic clean surfaces with Fe impurities protected from oxidation imbedded in the first crystallographic planes.

\begin{figure*}[t]
    \begin{center}
        \includegraphics[width=16 cm]{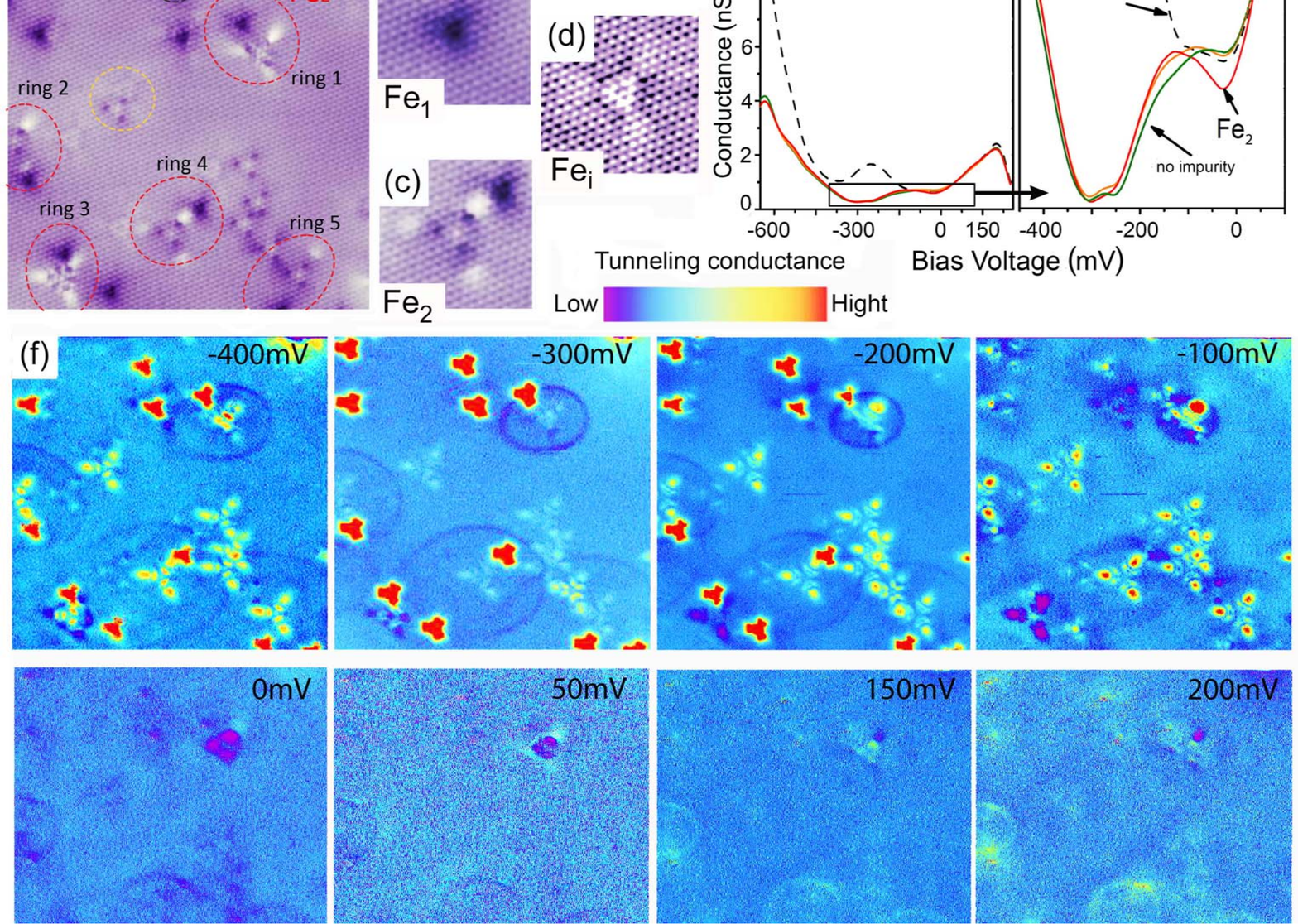}
           \caption{(a) STM topography image of Fe$_{0.2\%}$ Bi$_2$Se$_3$. V$_{sample}$=103~mV and I=177~pA. $26\times26 nm^2$. (b),(c),(d) - zoom on Fe$_1$, Fe$_2$ and Fe$_i$ , respectively. $6\times6 nm^2$. (e) spectral signatures of the defects recorded by STS on Fe$_1$, Fe$_2$ impurities and compared to the average spectrum. (f) Conductance images recorded at the same location than topography image (a). Stabilization parameters: V$_{sample}$=102.5 mV, I=180 pA. All STM/STS data were obtained at temperature T=1.5~K.}
      \label{Fig1}
        \end{center}
    \end{figure*}

The implanted Fe atoms produce at least three typical kinds of punctual defects shown in Fig.~1, each having a characteristic topographic fingerprint and a specific spectral signature in STM images.
Fe atoms may either occupy interstitial sites or substitute isovalent two distinct Bi sites at the subsurface \cite{song}. The surface plane of Bi$_2$Se$_3$ is composed of Se atoms so the location of Fe impurities in the lower atomic planes can be evidenced by simultaneous AFM-STM imaging: AFM imaging (see Fig.~2(a-b)) does not show any chemical contrast in the Se atomic lattice of the surface plane where the tunneling current reveals the presence of a triangular electronic
patterns due to the scattering of the electronic states on the buried impurity. In agreement with previous works we can identify these impurities in figure Fig.~1(a). Fe substitution at Bi sites in the second atomic layer from the surface are most often observed \cite{song}, Fig.1(b); these defects are denoted Fe$_{1}$ in the following. Other characteristic defects - Fe$_{2}$ - are associated with Fe substitution at Bi sites in the fourth atomic layer from the surface, Fig.~1(c) (for clearness, see Fig.~2(c)). In addition to these defects, a low concentration of Fe atoms occupying interstitial sites - Fe$_{i}$ - is also observed, Fig.~1(d).

\begin{figure}[h]
    \begin{center}
        \includegraphics[width=7 cm]{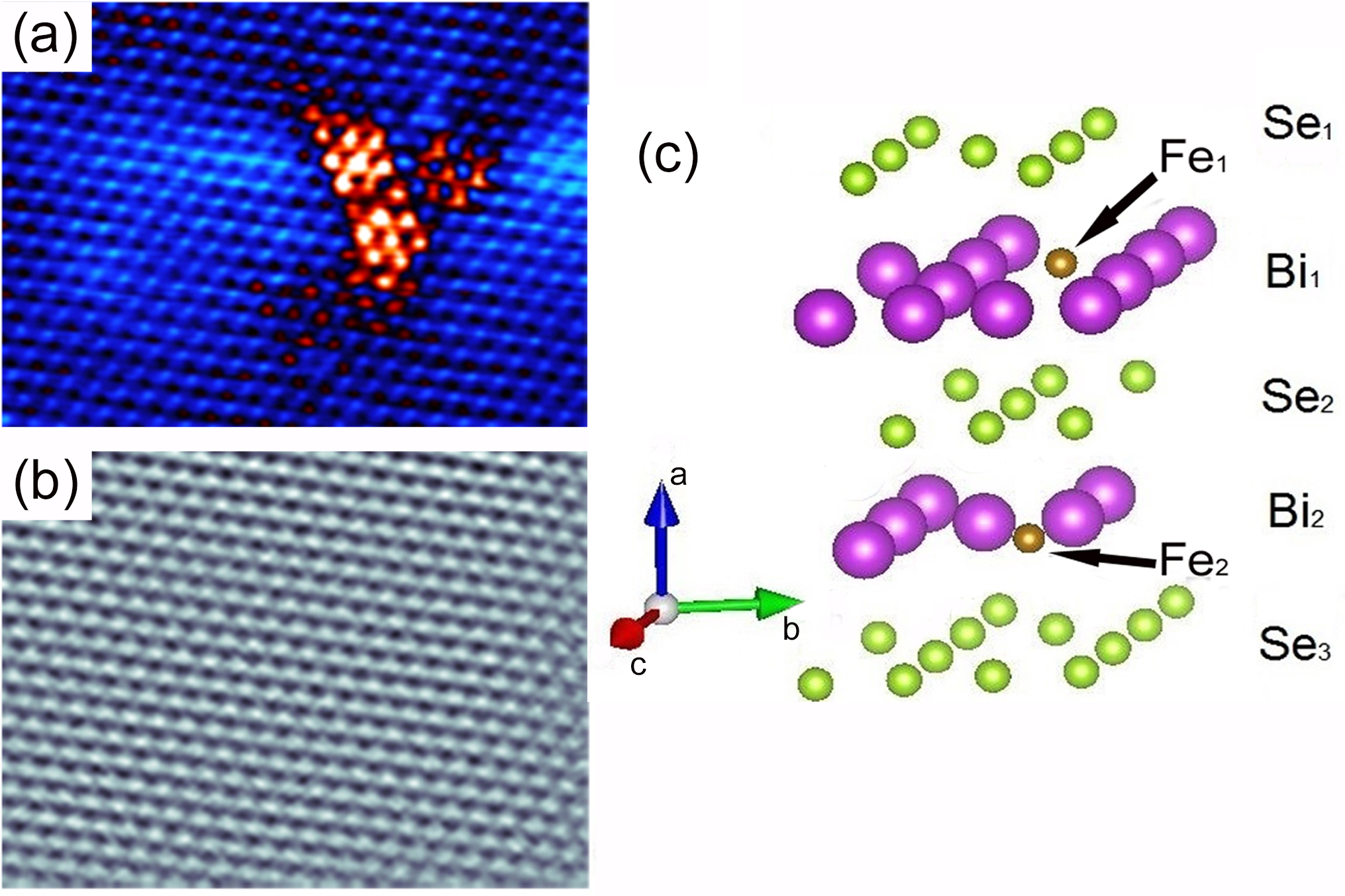}
           \caption{AFM-STM image of a buried impurity and crystal structure of Bi$_2$Se$_3$. (a)- current image; (b) - AFM simultaneous image. The feedback was made in AFM mode by keeping a constant frequency shift of 1Hz. (c) - two Fe substitutions for Bi sites are labeled Fe$_1$ and Fe$_2$, respectively.}
      \label{Fig1bis}
        \end{center}
    \end{figure}

A selection of $dI/dV(V)$ conductance images recorded by scanning tunneling spectroscopy and measured above and below the Fermi level $E^{s}_{F}$ is presented in Fig.~1(e). All these images were recorded in the area presented in topographic image Fig.~1(a).
In our experiments, Fe$_{1}$ defects exhibit a specific spectral signature at -300~mV presented in Fig.~1(e), giving rise to a strongly contrasted triangular shape in conductance images at negative voltages in Fig.~1(f). Fe$_2$ sites do not show any sharp signatures in the conductance spectra, Fig.~1(e), but they also exhibit regular patterns with triangular symmetry in the conductance images at negative voltages. Finally, Fe$_i$ sites only weakly perturb the electronic properties.

\begin{figure}[h]
    \begin{center}
    \includegraphics[width=7 cm]{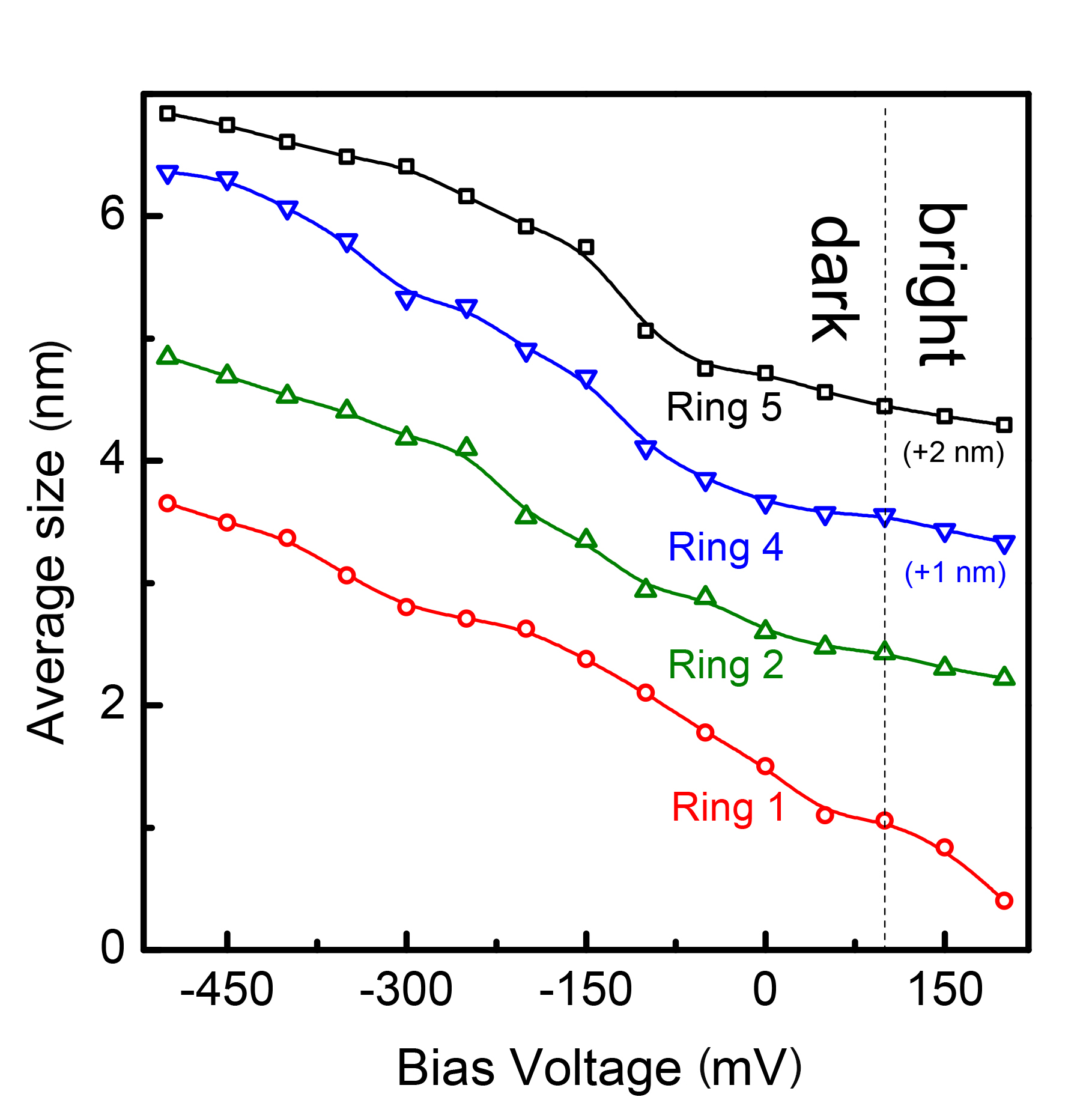}
    \caption{Evolution of the radius of ionization rings as a function of sample bias: $\sqrt{ab}/\pi$, where a and b are semi-axes of the elliptic rings. The spectral contrast switches from dark (at negative sample biases) to bright at approximately +100mV. For clarity +1 and +2 indicate that the average size is increased by +1 and +2 nm.}
      \label{Fig2}
        \end{center}
    \end{figure}

Circular patterns due to charging of Fe impurities are also visible in conductance images. The rings are seen to appear only around some specific defects and some defects are not related to charging rings. The ring contrast is dark for bias voltages ranging from -500~mV to +100~mV applied to the sample with respect to the tip. On the contrary, from +100~mV to +250~mV the rings appear brighter.
It has been shown in prior works that the observed rings can be unambiguously associated with the charging/discharging of specific impurities, triggered by the STM tip. The density of the conduction states being low, the presence of a metallic tip at biased voltage near the surface results in a local band bending in the sample below the tip. For a better understanding of this phenomenon, one can evoke a certain analogy between tip-induced band bending and the working principle of a MOS-FET: the biased tip and the vacuum barrier of the STM junction play the same respective roles as the metallic gate and the oxide layer in the FET. The band bending depends on the bias voltage of the tip with respect to the sample and also on the position of the tip. At certain defect-tip separation, the tip-induced band bending at the position of the defect can become significant enough to trigger its ionization. The Coulomb field of the ionized impurity produces an additional band shifting around a dopant. Consequently, the electron tunneling conditions through the STM junction are modified, resulting in circular features around dopants in tunneling conductance images. The imperfect circular symmetry of spectral features is mainly imposed by the symmetry of the electrostatic stray field of the imperfect conical tip.

Fe dopants are expected to exhibit two charge states in Bi$_2$Se$_3$: Fe$^{3+}$ and Fe$^{2+}$.
Fe$^{2+}$ state is more energetically favorable for Fermi level $E^{s}_{F}$ located closer to the conduction band minimum, while Fe$^{3+}$ dominates\cite{zhang} for lower $E^{s}_{F}$.
In the first case, the ionization of Fe$^{2+}$ into Fe$^{3+}$  should occur under
the upward band bending (sample voltages exceed flat-band voltage $V_{FB}$), while the ionization ring radius grows with increase of the sample voltage. In the second case, the ionization of Fe$^{3+}$ into Fe$^{2+}$ is possible under the downward band bending (sample voltage is lower than the flat-band voltage $V_{FB}$), while the ionization ring radius must grow with decrease of the sample voltage. The latter scenario is consistent with our observations of the dependence of ionization rings radii on the sample voltage (Fig.~3) and with previously reported results \cite{song} where Fe dopants were considered in the ${3+}$ state in the absence of STM tip. Thus, we conclude that the ionization rings observed in our experiments are associated with tip-induced switching of the dopant state from Fe$^{3+}$ into Fe$^{2+}$. Details regarding mechanisms of Fe impurity ionization under tip-induced band bending can be found in the supplementary material.

Although the basic physics behind our observation seems established \cite{song,zhang,west,schlenk}, we discovered two remarkable features which were not evidenced in prior works: (i) not all the defects exhibit charging rings in their vicinity (ii) the charging rings are revealed both at positive and negative voltages in our study and they manifest a puzzling contrast reversal.
In order to understand these new features, we first analyze the local environment of the charging rings. A closer inspection reveals that the rings are observed only around a very peculiar type of defects which have a composite nature. Individual Fe atoms do not provoke ionization rings in our sample. It is seen from STS images that the charging rings are always enclosing two impurities one being of Fe$_{2}$ nature and the other one of Fe$_{1}$ kind. These defects are separated by a few interatomic distances and form a doublet. The origin of this phenomenon seems to come from the position of $E^{s}_{F}$.

From our ARPES experiments (Fig.~4) we have established that $E^{s}_{F}$ of our heavily Fe doped samples (Fe$_{0.2\%}$Bi$_2$Se$_3$) is located in the bulk conduction band, i.e. at higher energy than that of the previously studied samples. The position of $E^{s}_{F}$ can be also lifted by native defects~\cite{suh}. A high $E^{s}_{F}$ should yield single defects to be in Fe$^{2+}$ state~\cite{zhang}. Thus, our Fe doped samples have to intrinsically exhibit Fe$^{2+}$ sites prior to any STM experiment.
Therefore, once the tip is brought towards the sample surface during a STM experiments, a downward tip-induced band bending will not result in any ionization of a single defect irrespective of its nature (Fe$_{1}$, Fe$_{2}$ or Fe$_{i}$). On the contrary, upward band bending would result in an ionization of the defects. We believe that such an ionization from Fe$^{2+}$ into Fe$^{3+}$ was not observed in our experiments since the highest sample voltage applied, +250~mV, was too low, while $V_{FB}$ in our experiment could be positive. Although an extraction of  a precise value of $V_{FB}$ is known to be a complicated task, simple considerations show that its sign in our experiments might be highly sensitive to the shape and sizes of our tip, since the work function for a platinum-iridium alloy of the tip is nearly the same as the electron affinity of Bi$_2$Se$_3$ equal to 5.3 eV \cite{affinity}. Under positive $V_{FB}$, bands are bend downward by the tip already at zero voltage. This could be a reason why the ionization from Fe$^{2+}$ into Fe$^{3+}$ was not observed.

\begin{figure}[h]
    \begin{center}
    \includegraphics[width=8 cm]{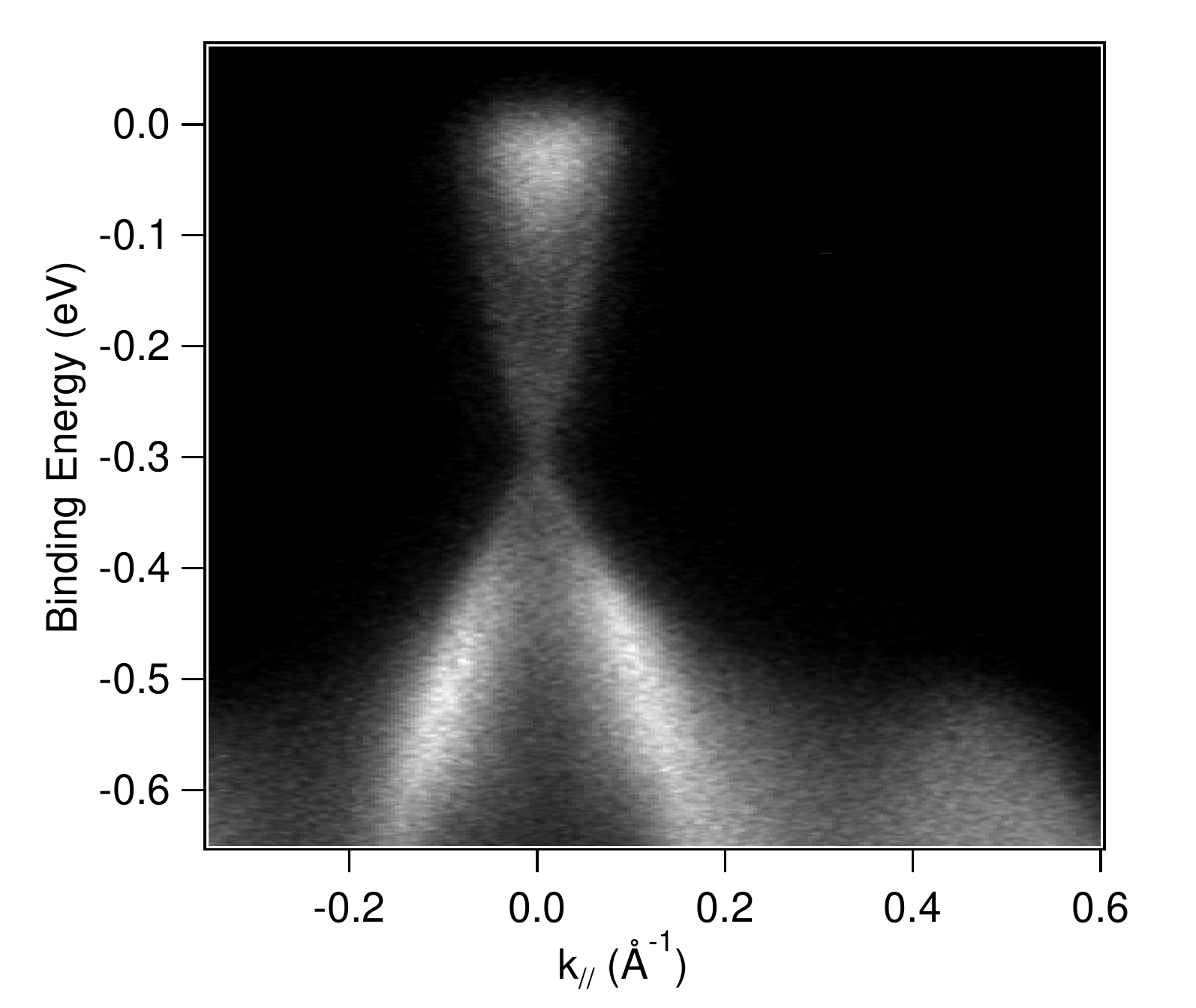}
    \caption{ARPES intensity image of Fe$_{0.2}\%$ Bi$_2$Se$_3$ taken at T = 10 K along the $\bar\Gamma–\bar M$ direction with photons of 21.2 eV.}
     \label{Fig3a}
        \end{center}
    \end{figure}

However, our experimental data evidence that the presence of Fe$_{1}$ defect in the surrounding of F$e_{2}$ defect (see Fig.~1(f)) gives rise to a different situation leading to the tip-induced ionization of Fe$_{2}$ defect. Ionization ring radius grows with the decrease of sample voltage, which implies that Fe$_{2}$ defect in the doublet is intrinsically in Fe$^{3+}$ state.  Microscopic mechanism for such an influence of Fe$_{1}$ defect on the charge state of Fe$_{2}$ defect is not so clear and deserves a separate study, but it might be related to the local shift of the band structure around Fe$_{1}$ impurity. This possible mechanism is discussed further in the supplementary material.

Another argument in support of our picture comes from the analysis of the contrast of the ring as a function of bias voltage in conductance images, e.g. Fig.~1(c,f). Most of rings appear bright at positive sample voltages and dark at negative sample voltages. A similar behavior has been reported in paper \cite{muzychenko} (see also paper \cite{kempen}), where Co islands on InAs surfaces have been studied. In Co/InAs, $E^{s}_{F}$ is also above conduction band minimum, in analogy with our samples, so it is natural to assume that this feature plays an important role in changing rings contrast. Indeed, ionization from Fe$^{3+}$ into Fe$^{2+}$ increases number of filled states in the valence band, but decreases their number in conduction band, which results in the interplay between the two opposite contributions to the conductance. Details regarding this competition, as well as an analysis of a connection between rings sizes and STM tip shape, can be found in the supplementary material.

In summary, we characterized and locally manipulated the electronic states of Fe dopants in Bi$_2$Se$_3$ crystals. In our experiments, because of the high $E^{s}_{F}$ in
our sample (due to heavy doping and native defects), the presence of $Fe_{1}$ impurities has been shown to allow for the charging process under the scanning tunneling tip. $Fe_{1}$ defects modify the transition threshold from one charge state to another of $Fe_{2}$ impurities in the neighborhood, possibly by locally doping the system with holes. Thus the composite structure of the defect is essential for the charge manipulation. Our observations evidence that the specific impurity neighborhood and the position of the Fermi energy have to be taken into account when considering the electron scattering on the disorder in topological insulators. Our work opens an interesting opportunity of engineering the charge (and magnetic) state of dopants in topological insulators by adding specific impurities in   their surrounding favoring or blocking their ionization.

See supplementary material for the description of details on ionization mechanisms and for the analysis of a relation between STM tip shape and ionization rings characteristics.

Authors acknowledge useful discussions with D. Muzychenko, A. A. Ezhov, and S. V. Zaitsev-Zotov.
The work of V.S., D.S., D.B., W.P. was supported by grants of Ministry of Education and Science of the Russian Federation in the frame work of the Federal Target Program "Research and development in priority areas of science and technology complex of Russia for 2014-2020", Grant nos.
14Y.26.31.0007, The ARPES studies supported by RFBR 16-02-00727 and 17-52-50080, French authors: D.R.,S.P.,S.V.,H.A.,T.C. acknowledge ANR (SUPERSTRIPES, MISTRAL), D.R. and S.P. acknowledge C'NANO Ile-de-France, DIM NanoK, for the support of the Nanospecs project, The VS gratefully acknowledge the financial support of the Ministry of Education and Science of the Russian Federation in the framework of  Increase Competitiveness Program of NUST «MISiS» (No.K3-2017-042), implemented by a governmental decree dated 16 th of March 2013, N 211. We thanks for partial support by the Program of Competitive Growth of Kazan Federal University.

\section{Supplementary Material}

\begin{figure*}[t]
	\includegraphics[width=0.4\linewidth]{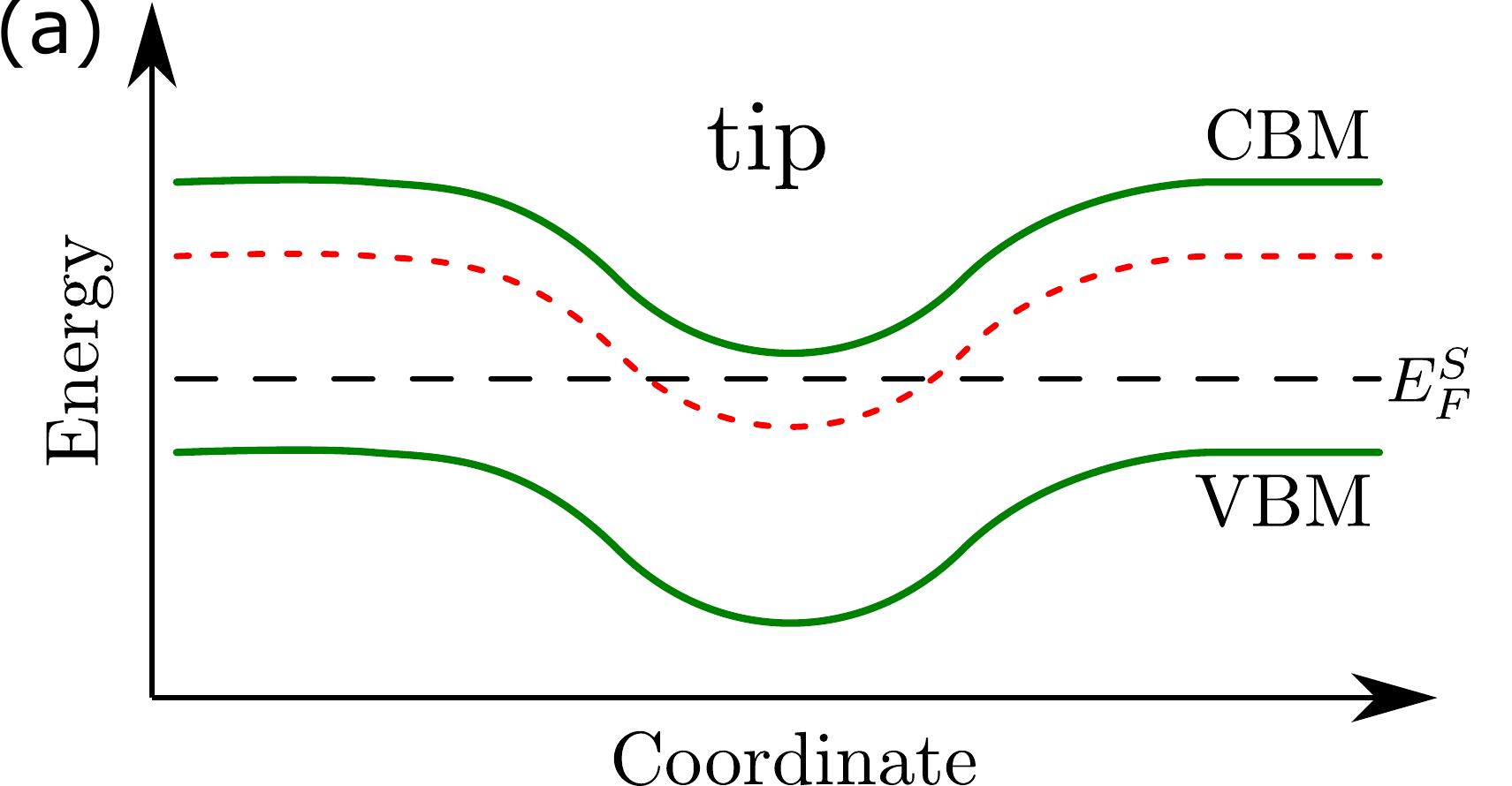} \includegraphics[width=0.4\linewidth]{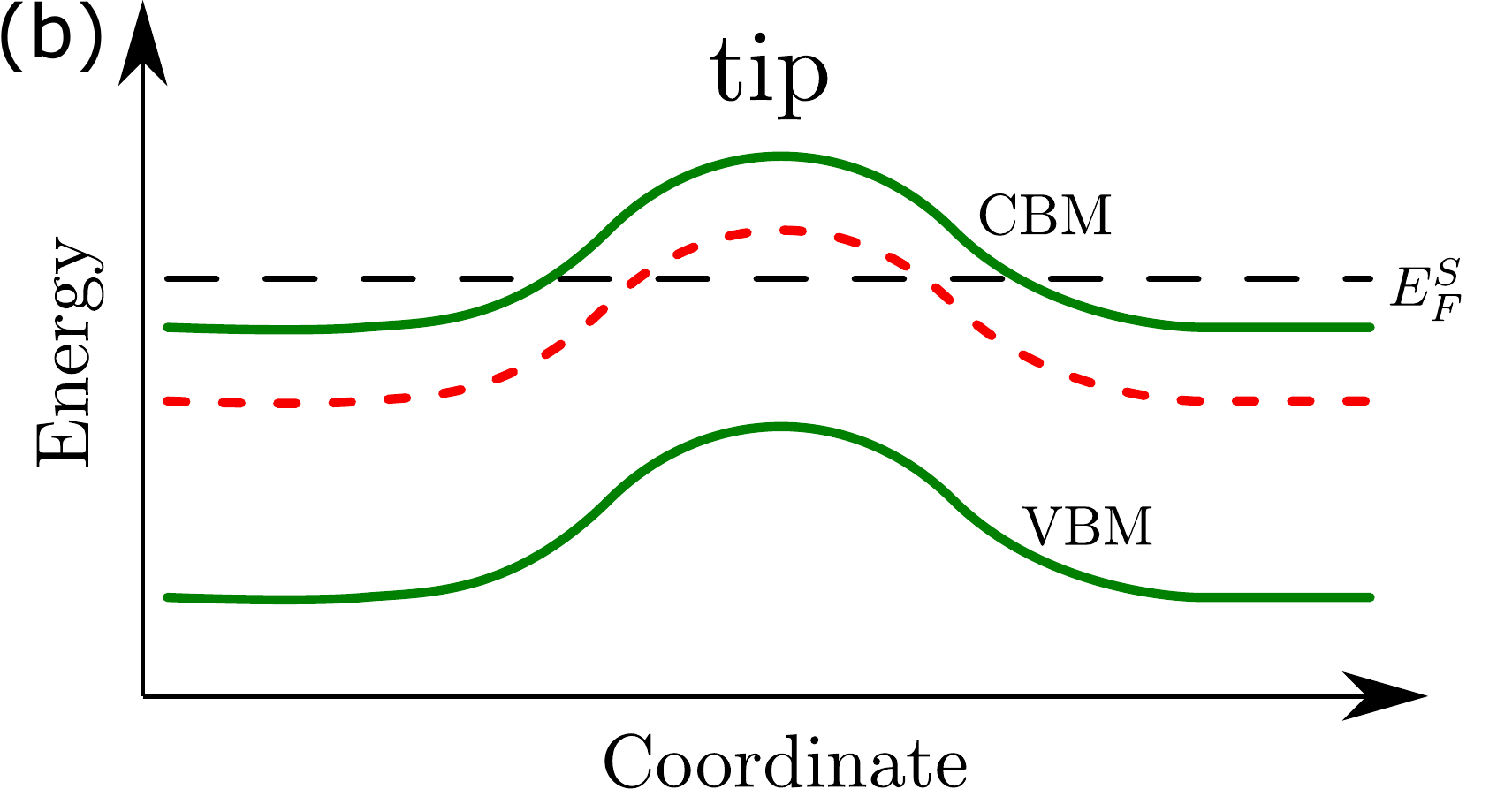}
\caption{(Color online) Schematic picture of the tip-induced band bending leading to the dopant ionization. Downward band bending (a) induced at sample voltages lower than $V_{FB}$ can ionize Fe impurity in Bi$_2$Se$_3$ from Fe$^{3+}$ into Fe$^{2+}$ provided $E^s_F$ (black dashed line) is below the charge transition level (red dotted line) in absence of the tip. Upward band bending (b) at voltages higher than $V_{FB}$ can ionize Fe impurity from Fe$^{2+}$ into Fe$^{3+}$ provided $E^s_F$ is above the charge transition level in absence of the tip.}
\label{S1}
\end{figure*}

\subsection{Ionization mechanism of a single Fe defect}
Fe impurity ionization in Bi$_2$Se$_3$ is linked to the sensitivity of its charge state to the position of $E^s_F$ with respect to the charge transition level (TL), the latter residing in the band gap \cite{zhang2013}. The tip-induced band bending leads to the local shift of the valence band maximum (VBM), conduction band minimum (CBM), and TL with respect to $E^s_F$. Charge state of the dopant thus changes provided TL comes in the resonance with $E^s_F$, which also implies that bands have to bend in appropriate direction (downward or upward bending depending on the mutual positions of TL and $E^s_F$). This is illustrated in Fig. \ref{S1}.
Note that slightly different interpretation of the same phenomenon was suggested prior \cite{song2012} in terms of the dopant level residing in the band gap. Within this picture \cite{song2012}, switching from Fe$^{3+}$  into Fe$^{2+}$ occurs when this level crosses VBM. However, it is easy to realize that both interpretations provide similar dependencies of ring radius as functions of energy: radius increases with decrease of bias sample voltage.

\subsection{Charge state of double defect}
From our ARPES experiment (Fig. 4) it is deduced that CBM is located nearly 200 mV below $E^s_F$. The spectra in Fig. 1(e) show that the conductance is strongly modified in this range of applied sample voltages near the Fe$_1$ defect compared to defect-free surface regions. To a certain extent, an influence of Fe$_1$ might be interpreted through the local doping effect of the system with an excess of holes, which bends the bands upward. The signatures of such an effect can be seen in Fig. 1(e). Indeed, for surface regions far away from Fe dopants, conductance as a function of voltage rises at voltages 200 mV below $E^s_F$ (position of CBM from our ARPES experiment). However, at Fe$_1$ defect, conductance becomes a decreasing function at the same voltages and it starts rising again at significantly higher voltages, which can be associated with shifted upwards position of CBM with respect to $E^s_F$. Let us stress that local hole doping at Fe$_1$ dopant was also revealed in Ref. [2] from the analysis of conductance spectra although they are not completely identical to our spectra in Fig. 1(e). The resulting band bending around the Fe$_1$ atoms may shift transition level closer to $E^s_F$ and make the crossover easier from 2+ to 3+ state of surrounding Fe$_2$ atoms. A described mechanism is illustrated schematically in Fig. \ref{S2} (a).
Our speculations are also consistent with the experimental fact that Fe$_2$ impurity in the double defect does not provoke Fe$_1$ atom to switch. Such an asymmetry between Fe$_1$ and Fe$_2$ impurities can be attributed to the surface band bending effect arising due to defects and dangling bonds \cite{suh2014}. Surface band bending implies that the position of $E^s_F$ with respect to the VBM is higher for Fe$_1$ compared to Fe$_2$, since the surface-defect distance is smaller for Fe$_1$. Thus, switching of Fe$_2$ impurity by Fe$_1$ is more likely than the opposite process.

\subsection{Dark and bright rings}
\begin{figure*}[t]
	\includegraphics[width=0.4\linewidth]{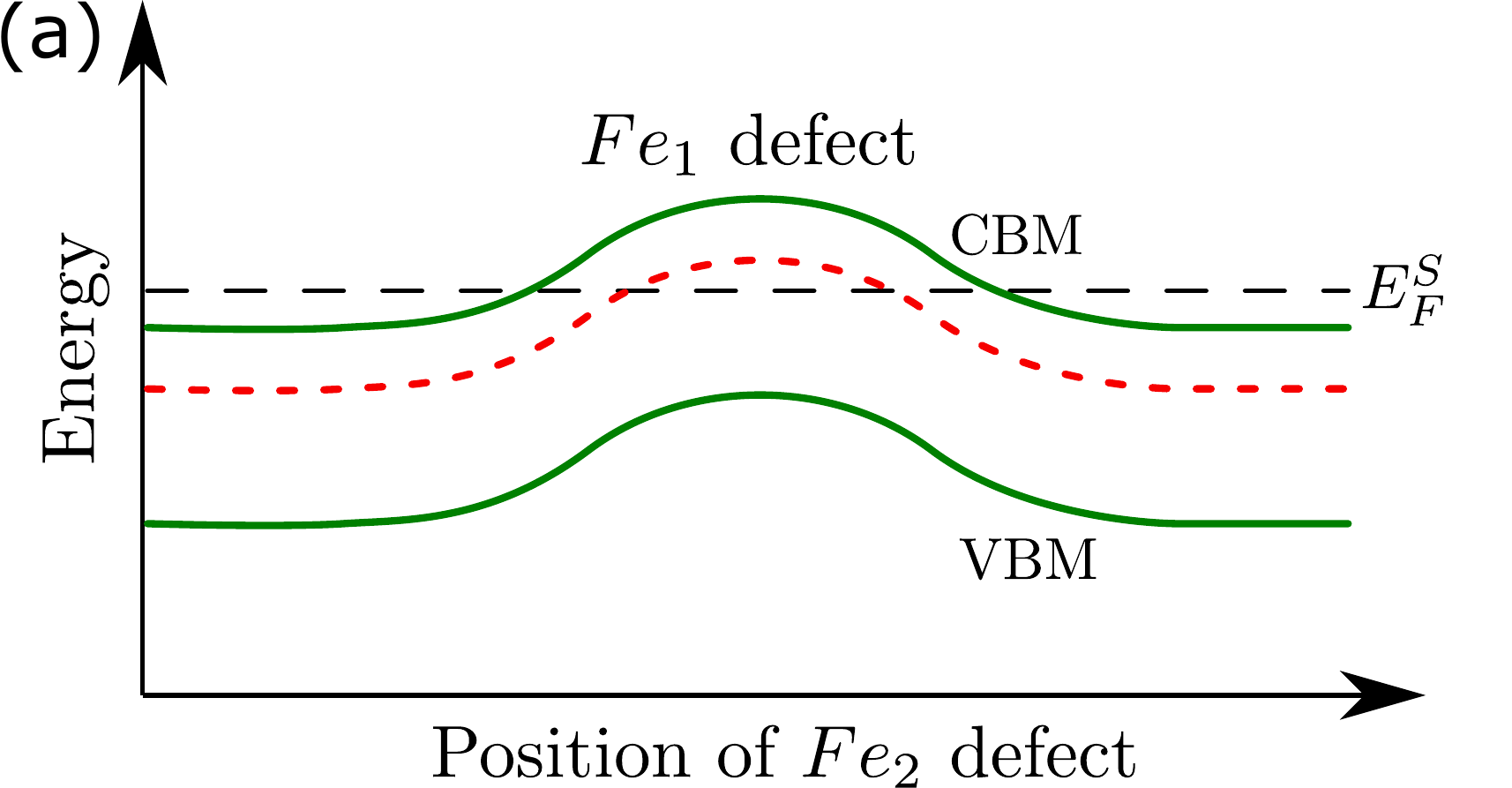} \includegraphics[width=0.4\linewidth]{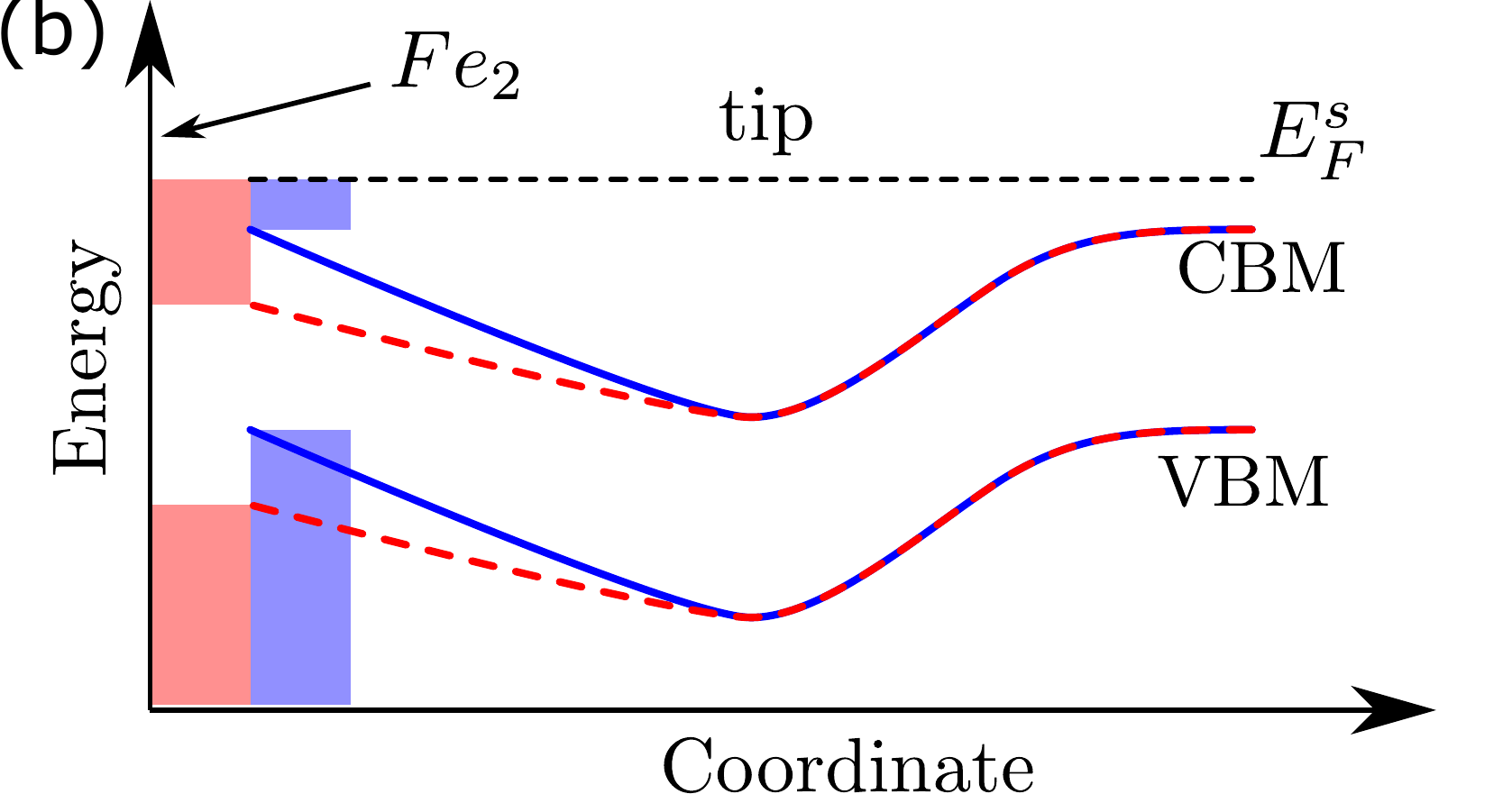}
	\caption{ (Color online) Schematic picture for the ionization of the double defect. Fe$_1$ locally changes environment for Fe$_2$ defect. A possible scenario is based on local hole doping effect (a), which makes Fe$^{3+}$ state more energetically favorable for the Fe$_2$ defect in the vicinity of Fe$_1$ defect - a transition level (red dotted line) is shifted above $E^s_F$ (black dashed line). The ionization of Fe$^{3+}$ state into Fe$^{2+}$, induced by the tip, lifts bands around Fe$_2$ defect (b) - solid blue lines correspond to Fe$^{2+}$ state, while dashed red lines refer to Fe$^{3+}$ state. Number of filled states, which belong to the valence band, increases locally, while the number of such states in the conduction band decreases, as illustrated by shaded areas.} \label{S2}
\end{figure*}

For samples with $E^s_F$ residing between VBM and CBM, one would expect the universal appearance of bright rings, which is consistent with prior experimental data \cite{song2012}. Indeed, the ionization from Fe$^{3+}$ to Fe$^{2+}$ results in the local band lifting, which increases a number of filled states in the valence band available for the electron tunneling to the tip \cite{song2012}. However, if underneath the tip CBM is below $E^s_F$ (downward or weak upward tip-induced band bending), such a lifting decreases number of filled electron states in conduction band in the area between the defect and the tip, thereby decreasing number of filled states available for electron tunneling to the tip.
This picture is illustrated schematically in Fig. \ref{S2} (b). Since the second contribution is more pronounced at downward tip-induced band bending, rings could appear dark at negative sample voltages and turn bright at positive sample voltages, when standard mechanism dominates.

\subsection{Ionization rings and tip shape}
STM tip   induces electrostatic potential $\phi(x,y,z)$ in the outer space.  A
precise shape of the tip of our experiments is unknown. However, the radius of curvature of the tip apex after cutting
of PtIr wire is expected to be much larger than the typical ring size. A distance between the tip termination and
Bi$_2$Se$_3$ surface is several nanometers. Thus, a tip radius is much larger than both the tip-surface separation and ring
radii. Hence, the problem of the field distribution can be treated as quasi one-dimensional.  It is known \cite{song2012}
that a carrier density of topologically protected gapless states is small and in the context of this problem the material
is considered as a bulk semiconductor. Free carriers   screen  the electrostatic potential inside the Bi$_2$Se$_3$ material which results in a decay of $\phi$ with the distance $z$. This results in a formation of a depletion layer in an area below the tip with the local  shift  of the energy bands \cite{Feenstra}. The screening of $\phi$ into the depletion layer defines the relation between the radius of the ring and tip voltage.  The two typical scenarios of screening are possible  depending on whether the Fermi level resides in (i) dopant levels or   (ii) in the conduction band. For a weakly doped semiconductor the scenario (i)  holds. In contrast,  according to ARPES data, our material is characterized by the regime (ii) and has similarities with a degenerate semiconductor. Decay of the electrostatic potential is dictated by the Poisson equation

\begin{equation}
\renewcommand{\theequation}{S\arabic{equation}}
\nabla^2 \phi(x,y,z)=\frac{e}{\varepsilon\varepsilon_0}\rho[\phi(x,y,z)] \label{poisson-0}
\label{eq1}
\end{equation}

where $\rho[\phi(x,y,z)]$   is the charge density   in the depletion layer.
Assuming a large tip radius, we neglect $x,y$ gradients of $\phi$ in (S1) and obtain

\begin{equation}
\renewcommand{\theequation}{S\arabic{equation}}
\partial_z^2 \phi(x,y,z)=\frac{e}{\varepsilon\varepsilon_0}\rho[\phi(x,y,z)]. \label{poisson-z}
\label{eq2}
\end{equation}

In semiconductor case (i) the r.h.s. $\rho[\phi(x,y,z)]=e\Delta n$ does not depend on $\phi$ where $\Delta n$ is carrier density. Within this approximation, the solution yields a quadratic decay of the potential $\phi(0<z<d)= \phi_0 (1-z/d)^2$. In the regime of degenerate semiconductor (ii) the relation  $\rho[\phi]=e \lambda \phi$ holds and  gives the exponential decay $\phi(z)=\phi_0 \exp(-z/\lambda)$. Nevertheless, for quasi-one-dimensional geometry, both dependencies for $\phi(z)$ result in the same relation between the tip shape and the dependence of the ring radius on the bias voltage up to the scaling factor along the $z$-axis. We parameterize this shape by $z$-coordinate $s(\bf r)$ of the point at the tip surface as a function of coordinate $\bf r = (x,y)$ in the plane of material surface with the origin underneath the tip termination. The two-dimensional coordinate $\bf r = (x,y)$ is associated with the ionization ring of a particular size. This ring, which is slightly deformed, is observed at certain bias voltage $V(\bf r)$. The tip shape is ultimately given by
\begin{equation}
\renewcommand{\theequation}{S\arabic{equation}}
s({\bf r}) \approx l \frac{V({\bf r})-V({\bf r}=0)}{V_0}. \label{s-v}
\label{eq3}
\end{equation}

\begin{figure}[h]
	\centering
	\includegraphics[width=0.99\linewidth]{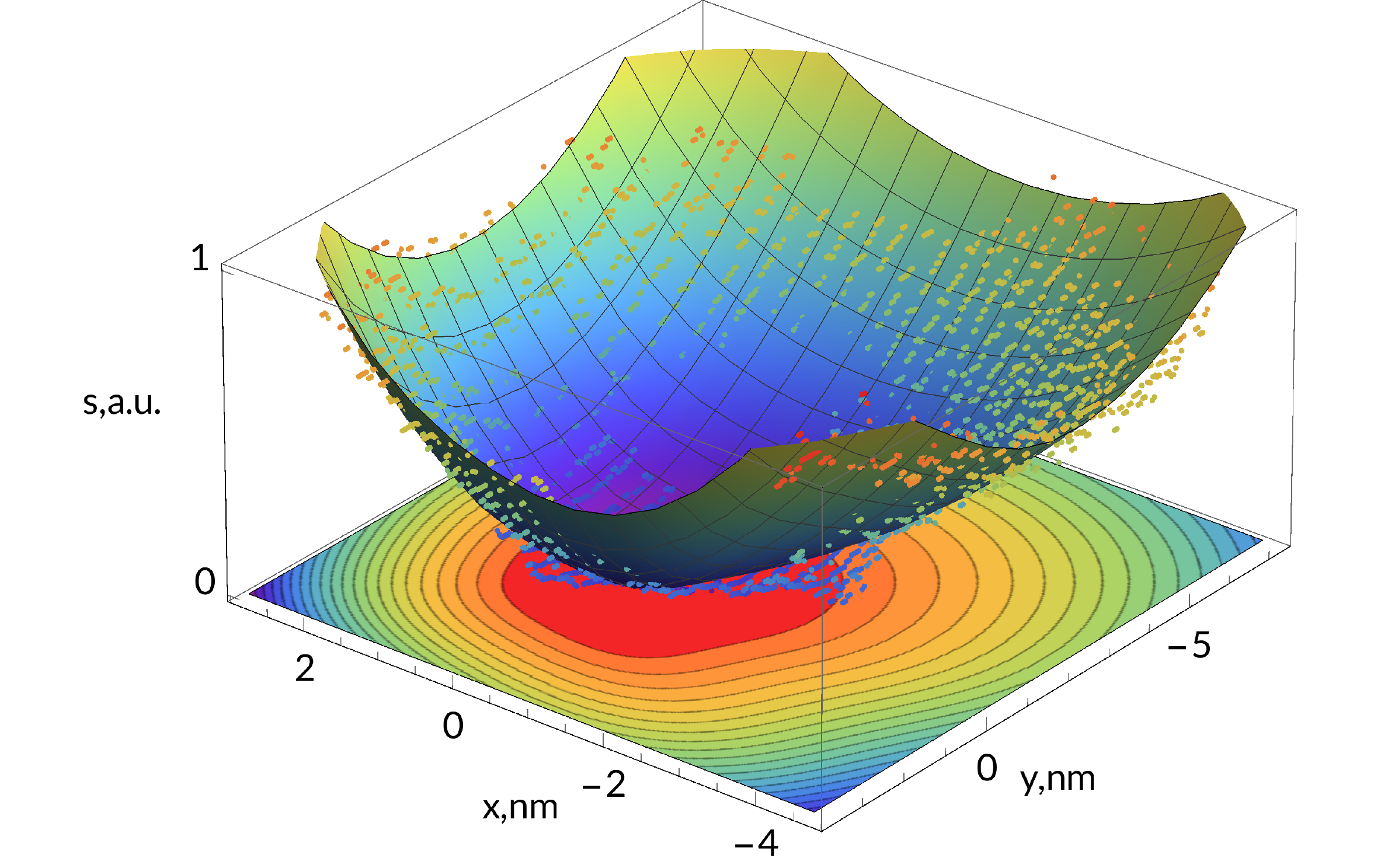}
	\caption{Reconstructed three-dimensional plot of the tip termination up to scaling factor in the vertical direction. The tip geometry $s(x,y)$ is obtained from the analysis of rings in STM maps.
	} \label{S3}
\end{figure}

This relation is quite universal and involves a couple of dimensional parameters $l $ and $V_0$  which are material specific. For instance, in the regime (ii), $l$ is given by the screening length $\lambda$ while $V_0$ is given by the difference between the local transition level and the Fermi energy (transition level is assumed to be above the Fermi energy for Fe$_{2}$ atom of the double defect). The existing  experimental data is not sufficient to determine both parameters. Thus,  Eq.~ (S3) can be used for the reconstruction of the tip geometry on a qualitative level, i.e., up to the unknown scaling in $z$-direction. The obtained tip shape is represented in Fig. \ref{S3} as a
three-dimensional reconstruction. Contours in the bottom of the three-dimensional plot are cross sections of the tip in the horizontal planes and they coincide with the same ring observed at different bias voltages. For the reconstruction, we used ring 1 from Fig. 1(a), because its size shrinks to zero in the investigated range of bias voltages (see Fig.~3), which is necessary to recover the shape of the tip termination. Since various rings exhibit similar dependencies of average size versus bias voltage, as depicted in Fig.~3, this procedure leads to similar tip shapes for different rings. The differences between positions of curves in Fig.~3 along the vertical axis seem to originate from local fluctuations of the transition level for Fe$_{2}$ atoms of double defects with respect to the Fermi energy.


\begin{thebibliography}{00}


\bibitem{hasan} M. Z. Hasan and C. L. Kane, Rev. Mod. Phys. {\bf 82}, 3045 (2010).

\bibitem{qi}X.-L. Qi and S.-C. Zhang, Rev. Mod. Phys. {\bf 83}, 1057 (2011).

\bibitem{fu98}L. Fu, C. L. Kane, and E. J. Mele, Phys. Rev. Lett. \textbf{98}, 106803 (2007).

\bibitem{moore}J. E. Moore and L. Balents, Phys. Rev. B \textbf{75}, 121306(R) (2007).

\bibitem{hsieh}D. Hsieh, Y. Xia, D. Qian, L. Wray, J. H. Dil, F. Meier, J. Osterwalder, L. Patthey, J. G. Checkelsky, N. P. Ong, A. V. Fedorov, H. Lin,
A. Bansil, D. Grauer, Y. S. Hor, R. J. Cava, and M. Z. Hasan, Nature {\bf 460}, 1101 (2009).

\bibitem{chen}Y. L. Chen, J.-H. Chu, J. G. Analytis, Z. K. Liu, K. Igarashi, H.-H. Kuo, X. L. Qi, S. K. Mo, R. G. Moore, D. H. Lu, M. Hashimoto,
T. Sasagawa, S. C. Zhang, I. R. Fisher, Z. Hussain, Z. X. Shen, Science {\bf 329}, 659 (2010).

\bibitem{cui}Y. Cui, N. Nilius, H.-J. Freund, S. Prada, L. Giordano, and G. Pacchioni, Phys. Rev. B \textbf{88}, 205421 (2013).

\bibitem{polyakov}A. Polyakov, H. L. Meyerheim, E. Daryl Crozier, R. A. Gordon, K. Mohseni, S. Roy, A. Ernst, M. G. Vergniory, X. Zubizarreta, M. M.
Otrokov, E. V. Chulkov, and J. Kirschner, Phys. Rev. B {\bf 92}, 045423 (2015).

\bibitem{yee}M. M. Yee, Z.-H. Zhu, A. Soumyanarayanan, Y. He, C.-L. Song, E. Pomjakushina, Z.Salman, A. Kanigel, K. Segawa, Y. Ando, and J. E. Hokman, Phys. Rev. B {\bf 91}, 161306 (2015).

\bibitem{eelbo}T. Eelbo, M. Wasniowska, M. Sikora, M. Dobrzanski, A. Kozlowski, A. Pulkin, G. Autes, I. Miotkowski, O. V. Yazyev, and R. Wiesendanger, Phys. Rev. B {\bf 89}, 104424 (2014).

\bibitem{xu}S.-Y. Xu, M. Neupane, C. Liu, D. Zhang, A. Richardella, L. A. Wray, N. Alidoust, M. Leandersson, T. Balasubramanian, J. Sanchez-Barriga, O. Rader, G. Landolt, B. Slomski, J. H. Dil, J. Osterwalder, T.-R. Chang, H.-T. Jeng, H. Lin, A. Bansil, N. Samarth, and M. Z. Hasan, Nat. Phys. \textbf{8}, 616 (2012).

\bibitem{hor}Y. S. Hor, P. Roushan, H. Beidenkopf, J. Seo, D. Qu, J. G. Checkelsky, L. A. Wray, D. Hsieh, Y. Xia, S.-Y. Xu, D. Qian, M. Z. Hasan, N. P. Ong, A. Yazdani, and R. J. Cava, Phys. Rev. B {\bf 81}, 195203 (2010).

\bibitem{wray}L. A. Wray, S. Y. Xu, Y. Xia, D. Hsieh, A. V. Fedorov, Y. San Hor, R. J. Cava, A. Bansil, H. Lin, and M. Z. Hasan, Nat. Phys. {\bf 7}, 32 (2010).

\bibitem{yu}R. Yu, W. Zhang, H. J. Zhang, S. C. Zhang, X. Dai, and Z. Fang, Science {\bf 329}, 61 (2010).

\bibitem{song}C.-L. Song, Y.-P. Jiang, Y.-L. Wang, Z. Li, L. Wang, K. He, X. Chen, X.-C. Ma, and Q.-K. Xue, Phys. Rev. B {\bf 86}, 045441 (2012).

 \bibitem{xiu}F. Xiu, L. He, Y. Wang, L. Cheng, L. T. Chang, M. Lang, G. Huang, X. F. Kou, Y. Zhou, X. W. Jiang, Z. G. Chen, J. Zou, A. Shailos, and K. L. Wang, Nat. Nanotech. {\bf 6}, 216 (2011).

 \bibitem{west}D. West, Y. Y. Sun, S. B. Zhang, T. Zhang, X. C. Ma, P. Cheng, Y. Y. Zhang, X. Chen, J. F. Jia, and Q. K. Xue, Phys. Rev. B {\bf 85}, 081305 (2012).

 \bibitem{schlenk}T. Schlenk, M. Bianchi, M. Koleini, A. Eich, O. Pietzsch, T. O. Wehling, T. Frauenheim, A. Balatsky, J.-L. Mi, B. B. Iversen, J. Wiebe,
A. A. Khajetoorians, Ph. Hofmann, and R. Wiesendanger, Phys. Rev. Lett. {\bf 110}, 126804 (2013).

\bibitem{gupta}D. H. Lee and J. A. Gupta, Nano Lett. \textbf{11}, 2004 (2011).

\bibitem{teichmann}K. Teichmann, M. Wenderoth, S. Loth, R. G. Ulbrich, J. K. Garleff, A. P. Wijnheijmer, and P. M. Koenraad, Phys. Rev. Lett. {\bf 101}, 076103 (2008).

\bibitem{wijnheijmer}A. P. Wijnheijmer, J. K. Garleff, K. Teichmann, M. Wenderoth, S.Loth, R. G. Ulbrich, P. A. Maksym, M. Roy, and P. M. Koenraad, Phys. Rev. Lett. {\bf 102}, 166101 (2009).

\bibitem{marczinowski}F. Marczinowski, J. Wiebe, J.-M. Tang, M. E. Flatt\'e, F. Meier, M. Morgenstern, and R. Wiesendanger, Phys. Rev. Lett. {\bf 99}, 157202 (2007).

\bibitem{graphene}V. W. Brar, R. Decker, H.-M. Solowan, Y. Wang, L. Maserati, K. T. Chan, H. Lee, C. O. Girit, A. Zettl, S. G. Louie, M. L. Cohen, and M. F. Crommie, Nat. Phys. \textbf{7}, 43 (2011).

\bibitem{muzychenko}D. A. Muzychenko, K. Schouteden, S. V. Savinov, N. S. Maslova, V. I. Panov, and C. Van Haesendonck, J. Nanosci. Nanotechnol. \textbf{9}, 4700 (2009).

\bibitem{Kapustin}A.A. Kapustin, V.S. Stolyarov, S.I. Bozhko, D.N. Borisenko, N.N. Kolesnikov, 2015, Zhurnal Eksperimental'noi i Teoreticheskoi Fiziki,  \textbf{148}, 2, 321 (2015)

\bibitem{zhang}J.-M. Zhang, W. Ming, Z. Huang, G.-B. Liu, X. Kou, Y. Fan, K. L. Wang, Y. Yao, Phys. Rev. B \textbf{88}, 235131 (2013).

\bibitem{suh}J. Suh, D. Fu, X. Liu, J. K. Furdyna, K. M. Yu, W. Walukiewicz, and J. Wu, Phys. Rev. B \textbf{89}, 115307 (2014).

\bibitem{affinity}C. D. Spataru and F. L\'eonard, Phys. Rev. B \textbf{90}, 085115 (2014).

\bibitem{kempen}J. W. G. Wild\"oer,  A. J. A. van Roij, C. J. P. M. Harmans, and H. van Kempen, Phys. Rev. B \textbf{53}, 10695 (1996).
\end{thebibliography}

\begin{thebibliography}{<num>}

	\bibitem{zhang2013} J.-M. Zhang, W. Ming, Z. Huang, G.-B. Liu, X. Kou, Y. Fan, K. L. Wang, and Y. Yao, Phys. Rev. B \textbf{88}, 235131 (2013).
\bibitem{song2012} C.-L. Song, Y.-P. Jiang, Y.-L. Wang, Z. Li, L. Wang, K. He, X. Chen, X.-C. Ma, and Q.-K. Xue, Phys. Rev. B \textbf{86}, 045441 (2012).
\bibitem{suh2014} J. Suh, D. Fu, X. Liu, J. K. Furdyna, K. M. Yu, W. Walukiewicz, and J. Wu, Phys. Rev. B \textbf{89}, 115307 (2014).
	\bibitem{Feenstra}R. M. Feenstra, S. Gaan, G. Meyer, and K. H. Rieder, Phys. Rev. B  \textbf{71}, 125316 (2005).
\end{thebibliography}
\end{document}